\documentstyle[aps,multicol,graphicx]{revtex}

\input{tcilatex}

\begin{document}
\title{Dielectric signatures of lattice instabilities at $32K$ and $245K$ in $%
La_{2-y}Sr_{y}MO_{4+x}$ ($M=Cu,Ni$) Cuprates and Nickelates}
\author{P. V. Parimi$^{1}$, N. Hakim$^{1}$, F. C. Chou$^{2}$, S. W. Cheong$^{3}$ and
S. Sridhar$^{1}$}
\address{$^{1}$Physics Department, Northeastern University, 360 Huntington Avenue,
Boston, MA 02115\\
$^{2}$Center for Material Science and Engineering, MIT, Cambridge, MA 02139\\
$^{3}$Physics Department, Rutgers University, Piscataway, NJ 08854}
\maketitle

\begin{abstract}
New dielectric transitions are observed at common temperatures $32K$\ and $%
245K,$\ in isostructural $La_{2}CuO_{4+x}$\ and $La_{5/3}Sr_{1/3}NiO_{4},$\
that are signatures of local lattice (octahedral) instabilities. The present
dielectric transitions reveal new aspects of the phase diagram of the
perovskite cuprates and nickelates. They suggest that competition and
coexistence of superconductivity with dielectricity occurs that is analogous
to that between superconductivity and anti-ferromagnetism. These results
also indicate that inhomogeneous electronic states, such as charge stripes
and oxygen ordering, are strongly connected to underlying lattice
instabilities.
\end{abstract}

\begin{multicols}{2}%

The presence of intrinsic electronic inhomogeneities in $%
La_{2-y}Sr_{y}MO_{4}:M=Cu,Ni$\ ($LSMO)$\ is now becoming widely recognized 
\cite{kastner98,xiong96,tranquada95,hammel90,hundley90}. Static stripe
phases have been observed in $La_{2-y}(Nd,Sr)_{y}CuO_{4}$\ \cite{tranquada95}%
, and in $La_{2-y}Sr_{y}NiO_{4}$\ \cite{lee97} with onset around $245K$.
Mechanisms suggested have been purely Coulombic (electron-electron) or
magnetic in nature with a strictly 2D planar picture ignoring the
involvement of lattice. In the $La_{2-y}Sr_{y}CuO_{4}$ and $%
La_{2-y}Ba_{y}CuO_{4}$ the global lattice transitions, such as LTO, Pccn,
and LTT are well known to be associated with the tilts of the octahedra \cite
{jorgensen92}. Phase separation into oxygen rich ($x>0$) and poor ($x=0$)
regions around $245(\pm 5)K$\ and miscibility gap have been observed \cite
{tranquada,toru95,reyes93,kremer92,ishihara00}, leading to the coexistence
of both superconductivity and antiferromagnetism in the phase separated
regions, and formation of intercalated oxygen layers with staging.
Significant jumps are seen in thermal expansion at $T_{c}$ in various HTS%
\cite{meingast91,lang92}. A correlation is observed between the lattice
modes and electronic susceptibiltiy in $La_{2-y}Sr_{y}CuO_{4}$\cite
{mcqueeney01}. These results urge investigation of lattice-charge coupling
and the resultant lattice-charge phase diagram similar to purely magnetic
and structural phase diagrams that have been obtained from spin and lattice
probes\cite{kastner98,jorgensen92}.

In this Letter we report the observation of new dielectric transitions at
common temperatures $32K$\ and $245K,$\ in $La_{2}CuO_{4+x}$\ and $%
La_{5/3}Sr_{1/3}NiO_{4},$\ that are signatures of local lattice (octahedra)
instabilities. The present results, which reveal a striking commonality of
the perovskite cuprates and nickelates, connect a series of other
measurements indicating anomalies in the vicinity of $32K$\ and $245K$. This
connection is made here for the first time, and reveals new aspects of the
phase diagram of these isostructural oxides not recognized before. These
results suggest that inhomogeneous electronic states such as charge stripes
and oxygen ordering are strongly connected to the underlying lattice
instabilities rather than arising from purely magnetic or electronic
interactions.

A number of conventional local probes have been used to investigate
structural instabilities such as XAFS, NMR, NQR, Neutron diffraction and
Mossbauer spectroscopies. These techniques are powerful local probes\cite
{egami,mihai93}, however, their sensitivity is limited for probing small
displacements of atoms from centrosymmetric structures. The present work
utilizes precision dielectric measurements, carried out using a highly
sensitive superconducting (Nb) microwave cavity, which reveal signatures of
subtle structural instabilities. The very high quality factor, $Q\simeq
10^{8},$ of the cavity makes this technique a powerful tool to investigate
small displacements of atoms from centrosymmetric structure especially in
non-magnetic insulators, where the polarization response dominates the spin
response due to the high frequency, as has been evidenced by strong
dielectric transitions in $YBa_{2}Cu_{3}O_{7-\delta }$\cite{zhai01}.

High quality single crystals of $La_{2}CuO_{4+x}$ $(x=0.0125,0.0175)$\cite
{blakeslee} and $La_{5/3}Sr_{1/3}NiO_{4}$\cite{lee97} were prepared by the
TSFZ method and  have been well characterized by a variety of other
techniques. The excess $O$ concentration, $x$ in $La_{2}CuO_{4+x}$, was
determined from dc magnetic SQUID susceptibility measurements and
well-established relation between the Neel temperature $T_{N}$ and $x$. The
sample is placed at the maximum of the microwave magnetic field $H_{\omega }$
of the $TE_{011}$ mode resonant at $10GHz$. The dielectric permittivity $%
\widetilde{\varepsilon }(T)=\varepsilon ^{\prime }(T)+i\varepsilon ^{\prime
\prime }(T)$ is determined from the measured parameters, shift in cavity
resonant frequency $\delta f(T)$ and resonance width $\Delta f(T)$\cite
{zhai00}. In all the measurements $\widetilde{\varepsilon }(T)$ is monitored
as the sample is warmed slowly at the rate of 1.3K/min.

The microwave dielectric permittivity $\widetilde{\varepsilon }_{a}(T)$ when
the microwave magnetic field $H_{\omega }\,\parallel a-axis$ of $%
La_{2}CuO_{4.0175}$ shows a clear transition at $32K$ and two additional
features whose onset is at $100K$ and $245K$ (Fig. 1). Below these
transition temperatures the change in $\varepsilon ^{\prime }(T)$ is
accompanied by loss peaks in $\varepsilon ^{\prime \prime }(T)$. The $32K$
transition is marked by a rapid increase in both $\varepsilon ^{\prime }(T)$
and $\varepsilon ^{\prime \prime }(T)$ as the temperature is decreased and
is strongly thermal history dependent. This transition is found to be
dependent on the cooling rate and annealing time at maximum temperature, $%
294K$ suggesting a glassy nature. A very strong transition is observed in $%
\widetilde{\varepsilon }(T)$ when the sample had been slowly cooled down
from $294K$ to $4K$ in about $16$ hours. However, rapid cooling of the
sample from 294K to 4K in 2hrs and subsequent measurement of $\widetilde{%
\varepsilon }(T)$ results in a weak transition. The strength of the
transition is dependent upon the previous maximum warm-up temperature $%
T_{max}$- thus this transition has features of thermal aging that are often
characteristic of dielectric transitions. An additional feature has been
observed at low $T=17K$, which is associated with incipient
superconductivity in micro domains as indicated by dc SQUID magnetization
measurement. This transition is suppressed when the sample is cooled rapidly.

Remarkably, $La_{5/3}Sr_{1/3}NiO_{4}$ also shows two dielectric transitions
for $H_{\omega }\,\parallel \,a$ in $\widetilde{\varepsilon }(T)$ at $32K$
and $245K$ as shown in Fig. 2. As in the case of $La_{2}CuO_{4.0175}$ the $%
32K$ transition in $La_{5/3}Sr_{1/3}NiO_{4}$ is suppressed when the sample
is quenched, indicating thermal aging, while the remainder of the data is
less sensitive to thermal history. It was also observed that the $32K$
transition is unaffected when the sample was annealed at and above 200K for
several hours. This indicates that the phase formed between $200K$ and room
temperature appears to be crucial in giving rise to the $32K$ transition.

The present data can be analyzed in terms of multiple dielectric modes, $%
\tilde{\varepsilon}(T)=\tilde{\varepsilon}_{\alpha }(T)+\tilde{\varepsilon}%
_{\beta }(T)+\tilde{\varepsilon}_{\gamma }(T)+...$, each of which is well
described by a Debye relaxation form with respect to the temperature
dependence

\[
\tilde{\varepsilon}(T)=\tilde{\varepsilon}_{\alpha }+\tilde{\varepsilon}%
_{\beta }+...=\sum_{i=\alpha ,\beta ,\gamma ,..}\frac{\varepsilon _{i0}(T)}{%
1-i\omega \tau _{i}(T)}, 
\]

where $\varepsilon _{i0}(T)$ are static dielectric functions and $\tau
_{i}(T)$ relaxation times for each mode.

In $La_{2}CuO_{4.0175}$ three contributions can be identified. $\tilde{%
\varepsilon}_{\alpha }(T)$ indicates the onset of a new dielectric mode
which turns on below 32K. We describe this mode with parameters $\varepsilon
_{\alpha 0}(T)=900(1-T/T_{d\alpha })$ with $T_{d\alpha }=32K$ and $\tau
_{\alpha }(T)=6.5\times 10^{-9}(\sec \cdot K)/T$. $\varepsilon _{\alpha
0}(T) $ is similar to an order parameter which grows below a transition. As $%
T$ is lowered, $\tilde{\varepsilon}_{\alpha }(T)$ increases initially due to
gradual displacement of Oxygen from centrosymmetric position leading to
growing polarization as described in the octahedra model latter. However
below a characteristic temperature $\tilde{\varepsilon}_{\alpha }(T)$ begins
to decrease because the dipoles are no longer able to follow the microwave
field. In $La_{5/3}Sr_{1/3}NiO_{4}$ the $32K$ dielectric mode has a much
weaker strength with $\varepsilon _{\alpha 0}(T)\sim 0.5(1-T/T_{d\alpha })$,
and requires a highly sensitivity measurement such as the present microwave
measurement. $\tilde{\varepsilon}_{\beta }(T)$ of $La_{2}CuO_{4.0175}$
associated with $T_{d\beta }=245K$ transition is described by $\varepsilon
_{\beta 0}(T)=10(1-T/T_{d\beta })$ with and $\tau _{\beta }(T)=1\times
10^{-9}(\sec \cdot K)/T.$ $\tilde{\varepsilon}_{\gamma }$(T) which is
dominant at higher T is described with a relaxation time $\tau _{\gamma
}(T)=8\times 10^{-13}\sec ^{-1}\exp (1000/T)$ characterized by an activation
energy $1000K$.

The present transitions are strikingly similar to the microwave dielectric
transitions at $60K$ and $110K$ observed in insulating $YBa_{2}Cu_{3}O_{6.0}$%
\cite{zhai01}, whose dielectric permittivity is also well described by eqn.
(1). These transitions were identified with lattice instabilities - change
in buckling angle for 110K transition - that have been observed in many
measurements, including ion channeling\cite{sharma00}.

It is important to stress that all these materials are $ABO_{3}$ type
perovskites, which are well known to show strong dielectricity, notable
examples being the ferroelectric $BaTiO_{3}$ and the quantum paraelectric $%
SrTiO_{3}$. The perovskites readily display a variety of competing
structural instabilities that lead to different ferroelectric states \cite
{muller}. In these materials it is well established that the dielectricity
(ferroelectricity) ensues from the strong polarizability of oxygen ion in
the presence of cations ($Ba,La,Sr,Ti,Cu.....$) and that the relevant
dynamical and critical features are controlled by a limited number of
parameters such as the effective coupling between the cation ($Ti,Cu$) and $%
O $. The presence of dipolar modes in $YBa_{2}Cu_{3}O_{7-\delta }$ arising
from coupling of the buckled $Cu-O$ planes with electric-diploe $Ba-O$
layers has been theoretically addressed by Shenoy, et al.\cite{Shenoy97}.

The surprising commonality of the dielectric transitions in both $%
La_{2}CuO_{4.0175}$ and $La_{5/3}Sr_{1/3}NiO_{4}$ strongly suggests
contributions from a common unit with a characteristic energy scale that in
these isostructural units is the $MO_{6}$ octahedron. Such a description
finds strong theoretical support from the work by Thomas\cite{noel90} where
it was shown that the dielectric response of $ABO_{3}$ type perovskites was
describable in terms of individual $BO_{6}$ octahedra. The dielectric
function of the present crystals is high and could arise from an induced
dipole moment extended over a cation-anion bond distance, in this case $M-O$
bond distance. Electron covalency between a cation and anion induces a large
dipole moment\cite{noel90,ishihara94} which originates from the charge
transfer between the anion and cation sites. In the $La_{2}MO_{4+x}$ and $%
La_{2-y}Sr_{y}MO_{4}$ the $MO_{6}$ octahedra are further coupled both
electronically and vibrationally, and form arrays of weakly coupled rigid
units. There are a certain number of very low frequency vibrational modes
(or rigid unit modes, RUMs\ ) of these units which do not distort the units
but flex the coupling between them. These RUMs are natural candidates for
the soft modes that typically drive displacive dielectric phase transitions%
\cite{giddy93}. Since the large inter layer strains between the $MO_{2}$ and 
$LaO$ planes\cite{goodenough90} can be relieved by tilting of the $MO_{6}$
octahedra or correlated ferrodistortive buckling of $MO_{2}$ plane with
displacement of $O$ atoms in the plane we attribute the $32K$ and $245K$
transitions to the onset of such a long wavelength, near- ferrodistortive
distortion. These displacements would result in a change in bond lengths and
thereby a change in effective $\tilde{\varepsilon}(T)$.

The suppression of $\varepsilon _{\alpha }(T)$ in the quenched samples of $%
La_{2}CuO_{4.0175}$ and $La_{5/3}Sr_{1/3}NiO_{4}$ is a manifestation of
competition between order and disorder. In the well annealed, slow cooled $%
La_{2}CuO_{4.0175}$ the excess oxygen forms intercalated layers as shown in
Fig. 3a and in $La_{5/3}Sr_{1/3}NiO_{4}$ the holes form charge ordered
stripes. However, rapid cooling does not result in ordering but highly
disordered oxygen as shown in Fig 3b. An intercalated oxygen atom dispalces
the apical oxygens and thereby leads to the distortion and tilt of the
octahedra. In the ordered state the majority of octahedra chains are
unperturbed by the excess-oxygen which leads to a strong transition. However
in the disordered state since the excess O are randomly distributed above or
below the octahedra plane as interstitial impurities they will have a
bearing on the entire plane. Consequently in the Debye model for the
disordered state $\tilde{\varepsilon}_{\alpha }(T)$ is well described by $%
\varepsilon _{\alpha 0}(T)=100(1-(T/T_{d\alpha })$ (solid line Fig.1) with $%
\tau _{\alpha }(T)$ being same as that of the ordered state for this mode.
The role of disorder in suppressing the $32K$ transition is evidenced in the
measurement on flux grown crystals of $La_{2}CuO_{4}$ (not shown) where our $%
10GHz$ dielectric data shows suppression of the $32K$ transition, consistent
with earlier dielectric measurements \cite{reagor,kastner98}.

In the case of $La_{5/3}Sr_{1/3}NiO_{4}$ suppression of $\varepsilon
_{\alpha }(T)$ indicates that high temperature ( $>$ $245K$ ) annealing
determines the strength of transition at $32K$. We further argue that charge
ordering per se has a driving lattice component in it because purely
electronic states are not expected to be dependent on thermal annealing. We
believe that quench disorder (rapid cooling) would lead to a change in the
structure of charge stripe formation which would affect the tilts of
octahedra at $32K$ resulting in a weak dielectric transition.

Another temperature scale that is common to $La_{2}CuO_{4+x}$ and $%
La_{5/3}Sr_{1/3}NiO_{4}$ is around $245K$ (see Figs.1 \& 2). The $245K$
transition in $\widetilde{\varepsilon }(T)$ of $La_{2}CuO_{4+x}$ is
attributed to oxygen ordering observed in neutron diffraction measurements
by Tranquada, et. al.\cite{tranquada}. In $La_{2-x}Sr_{x}NiO_{4}$, a
detailed elastic neutron diffraction study has demonstrated charge ordering
around $T_{cCO}=245K$ into domain walls or stripes\cite{lee97}. In the
dielectric response this leads to the onset of a dielectric mode associated
with the stripe formation (see Fig.2). In both these materials we can
associate a dielectric mode $\tilde{\varepsilon}_{\beta }(\omega ,T)$ with
the $245K$ feature. That Oxygen ordering and charge ordering take place at
the same temperature in $LCO$ and $LSNO$ is now understandable since it is
the underlying lattice instability, observed here as a dielectric transition
at $245K$, that drives these transitions in these two isostructural
compounds.

Observation of the $32K$ dielectric transition and its relation to the $%
MO_{6}$ octahedra modes also explains the origin of anomalies at this
temperature in NQR, thermal expansion and specific heat. Thermal expansion
measurements \cite{lang92} reveal changes at $32K$ and $36K$ in the
coefficient of thermal expansion indicating lattice changes. Migliori et al.%
\cite{migliori90} have observed a feature with onset at $32K$ in the
specific heat of $La_{2}CuO_{4}$. NQR measurements \cite{brom00} reveal a
change in the $^{139}$La spin-lattice relaxation rate and spin-spin
dephasing rate between $30K$ and $38K$ \cite{brom00}. Anelastic measurements
revealed softening of $La_{2}CuO_{4}$ lattice with onset around $32K$\cite
{cordero98}. Recently, Takao, et. al.\cite{wakimoto00} observed a change in
sound velocity at $36K$ in $La_{2}CuO_{4.05}$ which is found to be present
even after suppression of superconductivity by magnetic field. While all
these measurements indicate a structural change at $32K$ (and 36K) our
microwave dielectric measurements connect these changes to the instability
of the octahedra. The $MO_{6}$ octahedron is strongly coupled to the $La$
nuclear moment through an orbital hybridization mediated by apical oxygen 
\cite{yoshinari98}. Therefore, the subtle tilt/distortion of the octahedra
directly affects the $^{139}$La spin-lattice relaxation rate. Thus our
microwave measurements indicate that dipolar-phonon mediated mechanisms are
important and play a significant role in inducing many electronic
transitions in these materials.

The coexistence of , and competition between dielectricity and
superconductivity is clearly evident in the result obtained on a single
crystal $La_{2}CuO_{4}._{0125},$ for $H_{mw}\parallel a$ a dielectric
transition is observed at $31K$ followed by onset of superconductivity at $%
27K$ (inset Fig. 1b). This observation coupled with the results on $x=0.0175$
crystal (Fig. 1) suggests that the dielectric and superconducting
transitions are independent of each other, and when the crystal
superconducts the dielectric response is overshadowed below $T_{c}$. Its
important to note that the competition between ferroelectricity and
superconductivity is well-known in the A-15 compounds, and its possible
relevance to perovskite oxides was noted shortly after the discovery of high
temperature superconductivity \cite{Muller}.

The importance of the present results is emphasized in a new phase diagram
(Fig. 4) constructed from the present microwave observations and other
results reported earlier on $LMO$ and $LSMO$, which together reveal
unambiguously new structural instabilities at $32K$ and $245(\pm 5)K$. The
structural instability at $245(\pm 5)K$ leads to charge ordering in the case
of $LSNO$ and oxygen ordering in the case of $LCO$. The fact that these
compositions have different hole doping suggests that these transitions are
purely structure dependent and almost independent of hole doping in the low
doping regime. In contrast the magnetic ($T_{N}$) and superconducting
transitions ($T_{c}$) are well-known to be doping dependent. It is important
to note a striking similarity of the present transitions in $LSCO$ and $LSNO$
to multiple temperature scales observed in ion channeling measurements on $%
YBa_{2}Cu_{3}O_{7-\delta }$ \cite{sharma00}. The present observation of
signatures of octahedra instability at common temperatures in both
underdoped (no stripes) and doped (presence of stripes) regimes confirms
that charge stripes and oxygen ordering are coupled to the underlying
lattice instabilities rather than resulting purely from magnetic or
electronic interactions. There are a few other temperatures where lattice
anomalies were reported previously \cite{lang92}. However, not all the
lattice instabilities would effect the octahedra (RUMs) and correspondingly
have dielectric signatures. In the phase diagram we emphasized only those
results that show anomalies at $32K$ and $245K$ and are related to the
present dielectric transitions. The present dielectric transitions at common
temperatures underscore the importance of subtle local structural
instabilities that result in dramatic changes in the electronic properties
of perovskite oxides.

We thank R. S. Markiewicz and S. R. Shenoy for enlightening discussions.
This work was supported at Northeastern by the ONR and NSF, and at MIT by
NSF-DMR-9808941.

\narrowtext%

\begin{figure}
\caption{Microwave (10GHz) dielectric constant
 $\varepsilon^{\prime} (T)$ and  $\varepsilon ^{\prime \prime} (T) $ 
of $La_2 Cu O_{4.0175}$. Note the onset of dielectric modes at 
$T_{d\alpha} = 32K$ and $T_{d\beta} = 245K$, and the 
difference in the strength of dielectric function between slow cooling and rapid cooling.
Solid lines are fits to Debye ralaxation form. Inset (bottom panel) shows the $T_{d\alpha}$ 
transition in $La_2 CuO_{4.0125}$. }
\end{figure}
%

\begin{figure}
 \caption{Microwave (10GHz) dielectric constant $\varepsilon\prime (T) $ and
$\varepsilon\prime\prime (T) $
of $La_{5/3} Sr_{1/3} NiO_{4.0}$ (top panel). Inset (top panel) shows supression 
of 32K mode in the quenched sample. Note that the dielectric modes at 32K and 
245K are present
in both LCO and LSNO (bottom panel). }
\end{figure}
%

\begin{figure}
 \caption{ Schematic diagram of intercalated oxygen ordered (a) and disordered (b) lattice 
showing cross section of the rigid units of $MO_6$ (shaded squares) that lie in a plane.
M occupies center of each shaded square and O at its center. Dashed square represents 
$MO_2$ plane of one unit cell. Solid circles represent excess oxygen that lie below and above the
$MO_6$. The tilt pattern of the octahedra determines $\varepsilon (T)$. The distortion of the octahedra tilt pattern
 in the disordered phase explains the supression of 32K mode in quenched samples. }
\end{figure} %

\begin{figure}
\caption{ Phase diagram of LMO and LSMO. Two temperature scales 32K and 245K 
are found which are  characteristic of these isostructural systems.
Note that these two temperature scales are
independent of (small) hole doping.  Solid diamonds  represent present microwave 
dielectric transitions. Several other experimental results also
reveal anomalies at 32K and 245K including neutron diffraction $\bullet$ [8], 
sound velocity $\Diamond $[32], thermal expansion $\circ$ [14],  NMR $triangle$ [4],
thermoelectric power solid square [5], resistivity, dc magnetization square [34], 
and thermal expansion solid triangle [14]. 
The dashed lines representing $T_c$  and $ T_N$ are for $La_{2-y} Sr_y CuO_{4+x}$ system.   
}
\end{figure} %

\end{multicols}%

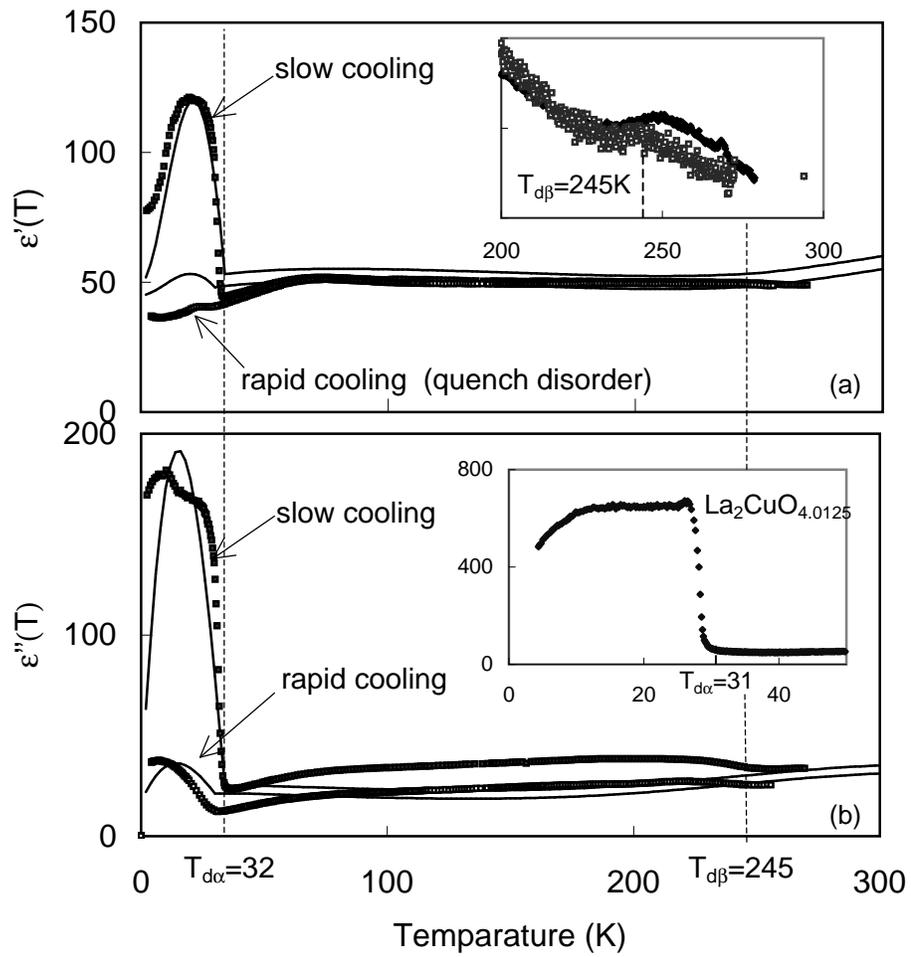
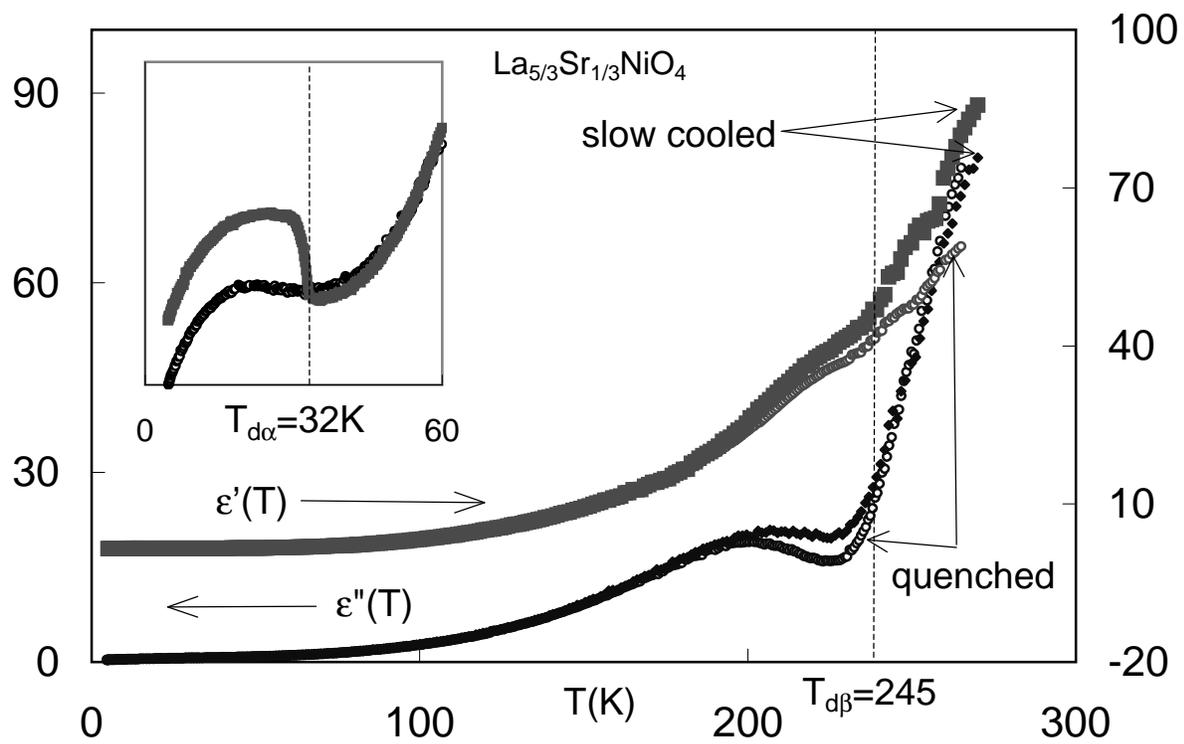
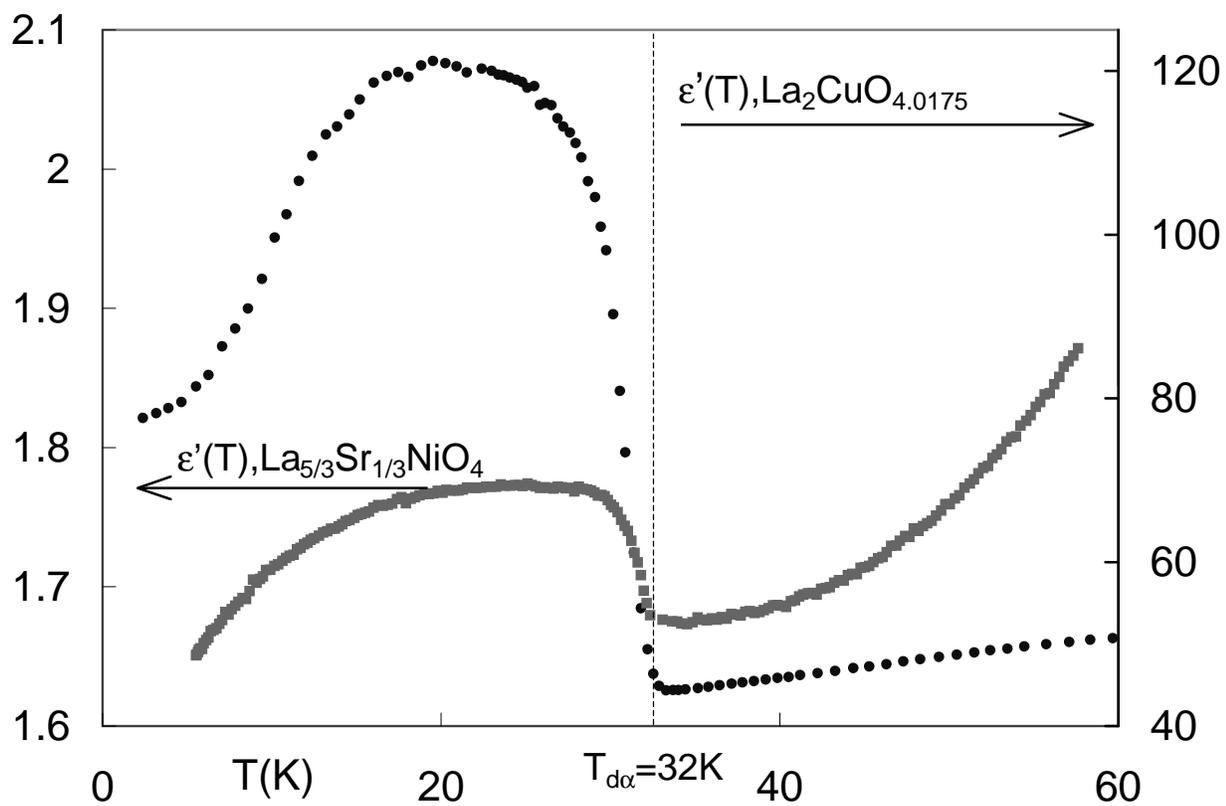
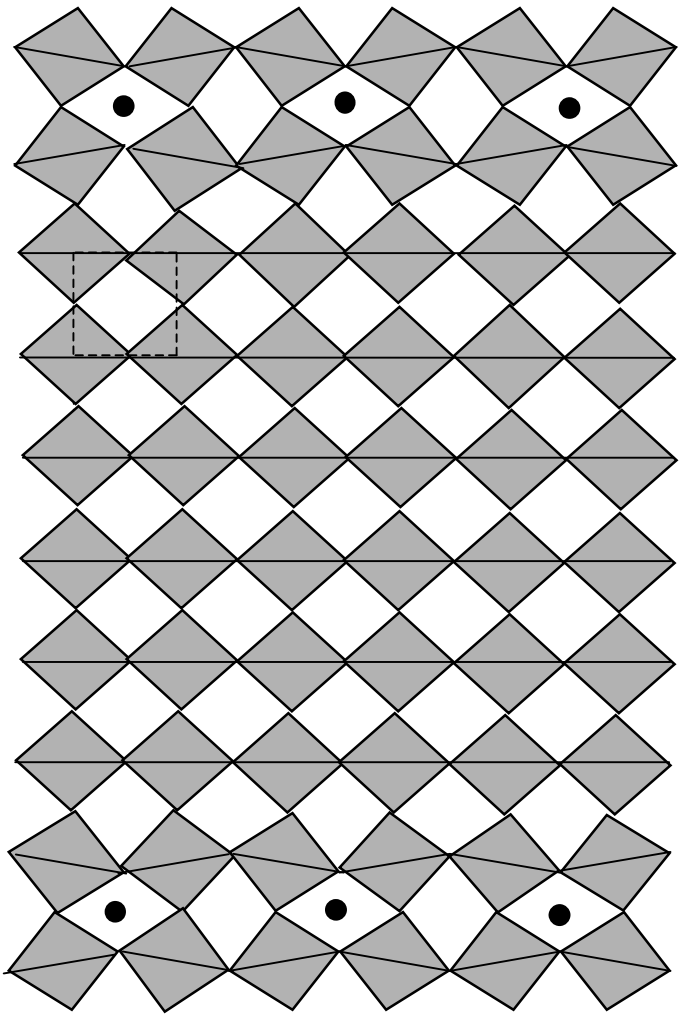
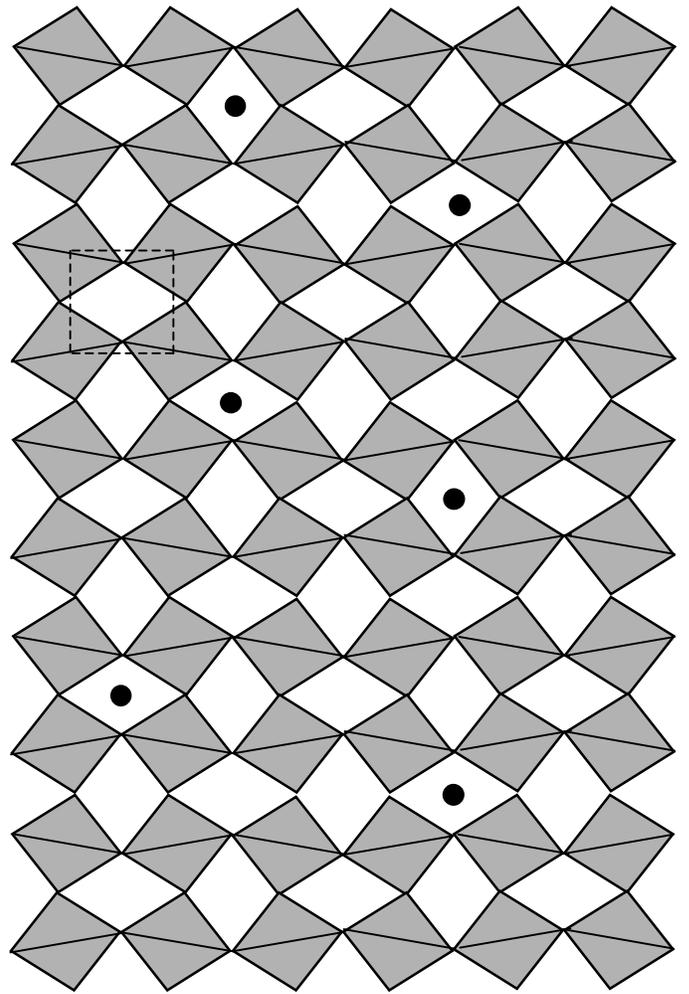
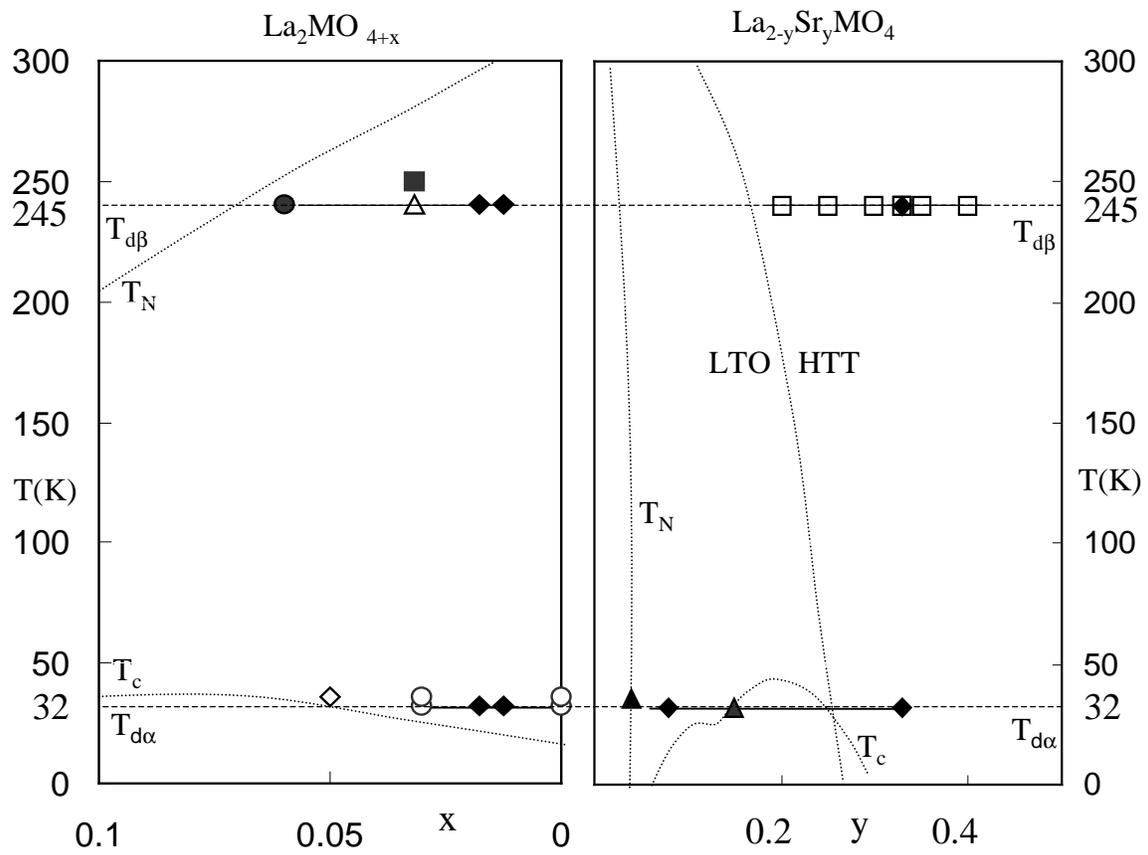
\end{document}